\documentclass[aps,twocolumn,amsmath]{revtex4}
\usepackage{graphicx}
\usepackage{array}
\usepackage{hhline}
\usepackage{longtable}

\begin{document}

\title{Superfluid Weight, Free Carrier Density, and Specific Heat

of the $d=3$\: $tJ$ Model at Finite Temperatures }
\author{Michael Hinczewski and A. Nihat Berker}
\affiliation{Department of Physics, Istanbul Technical University,
Maslak 34469, Istanbul, Turkey,} \affiliation{Department of Physics,
Massachusetts Institute of Technology, Cambridge, Massachusetts
02139, U.S.A.,} \affiliation{Feza G\"ursey Research Institute,
T\"UBITAK - Bosphorus University, \c{C}engelk\"oy 81220, Istanbul,
Turkey}
\begin{abstract}
The superfluid weight, free carrier density, and specific heat of
the three-dimensional $tJ$ model are calculated by
renormalization-group theory.  We find that optimal hole doping for
superfluidity occurs in the electron density range of $\langle n_i
\rangle \approx$ 0.63--0.68, where the superfluid weight
$n_s/m^\ast$ reaches a local maximum.  This density range is within
the novel $\tau$ phase, where the electron hopping strength
renormalizes to infinity, the system remains partially filled at all
length scales, and the electron-hopping expectation value remains
distinctively non-zero at all length scales.  The calculated
superfluid weight drops off sharply in the overdoped region.  Under
hole doping, the calculated density of free carriers increases until
optimal doping and remains approximately constant in the overdoped
region, as seen experimentally in high-$T_c$ materials. Furthermore,
from calculation of the specific heat coefficient $\gamma$, we see
clear evidence of a gap in the excitation spectrum for the $\tau$
phase.

\end{abstract}
\maketitle
\def\s{\rule{0in}{0.28in}}

\section{Introduction}

The variation of the superfluid number density $n_s$ with
temperature and carrier doping is of fundamental importance in
describing the unique properties of the superconducting state in
high-$T_c$ cuprates.  Experimentally, muon-spin-rotation techniques
are used to determine the closely related quantity $n_s/m^\ast$
(also known as the superfluid weight), where $m^\ast$ is the
effective mass of the carriers in the superfluid.  In the underdoped
region of high-$T_c$ materials, $n_s/m^\ast$ increases with doping,
and the low-temperature superfluid weight is correlated with
$T_c$.\cite{Uemura, Bernhard1}  As the materials are doped past the
optimal value (where $T_c$ is the highest), $n_s/m^\ast$ peaks and
rapidly decreases.\cite{Niedermayer, Locquet, Bernhard2}  The
decrease in $n_s/m^\ast$ is surprising since the total density of
free carriers saturates at optimal doping and remains almost
constant in the overdoped region.\cite{Puchkov} By contrast, in a
conventional superconductor, described by BCS theory, these two
quantities have the same doping dependence.

The $tJ$ model is a promising starting point in understanding these
properties of cuprate superconductors.  Mean-field $U(1)$ and
$SU(2)$ slave-boson theories of the $tJ$ Hamiltonian have reproduced
some aspects of the doping and temperature dependences of
$n_s/m^\ast$.\cite{LeeSalk1,LeeSalk2}  More direct, unbiased
numerical techniques applied to a $4\times 4$ $tJ$ cluster have
observed a large peak in $n_s/m^\ast$ in the same region where
pairing correlations indicate a superconducting ground
state.\cite{DagottoRiera}  A general limitation of these types of
studies is that no finite-cluster approach can unambiguously
identify phase transitions in the system, or exhibit the
non-analytic behavior of thermodynamic quantities at these
transitions.

Alternatively, the physics of the bulk model can be studied through
the position-space renormalization-group method, which has been used
to determine the phase structure and thermodynamic properties of the
$tJ$ and Hubbard models at finite
temperatures.\cite{FalicovBerkerT,FalicovBerker,Hubshort,Hubnew}  In
particular, Falicov and Berker's calculation for the $tJ$ model in
$d=3$ with the realistic coupling $J/t = 0.444$ produced a rich,
multicritical phase diagram~\cite{FalicovBerkerT, FalicovBerker},
with a novel low-temperature phase (called ``$\tau$'') for $30-40\%$
hole doping where the electron hopping strength in the Hamiltonian
renormalizes to infinity under repeated scale changes, while the
system remains partially filled. This is the possible signature of a
superconducting phase, and it is notable that a similar phase was
also observed in the $d=3$ Hubbard model.\cite{Hubshort,Hubnew}

Our present study further develops this renormalization-group
method, to yield the superfluid weight of the $tJ$ model as a
function of temperature and hole doping.  Our approach reproduces
phenomenological features of high-$T_c$ materials. In particular we
find that optimal doping is located in the vicinity of the $\tau$
phase, where $n_s/m^\ast$ peaks and then is sharply reduced with
overdoping. Moreover, we also find that the density of free carriers
increases until optimal doping, and saturates in overdoped region.
These results suggest that the $\tau$ phase might indeed correspond
to the superconducting phase in cuprates.  Further supporting this
idea, we present specific heat calculations that show clear evidence
of a gap in the quasiparticle spectrum for the $\tau$ phase.

\section{The $tJ$ Hamiltonian}

We consider a $d$-dimensional hypercubic lattice where the $tJ$
model for electron conduction is defined by the Hamiltonian
\begin{multline}
\label{eq:1} H = P \left[ \tilde{t} \sum_{\langle ij
\rangle,\sigma} \left(c^\dagger_{i\sigma}c_{j\sigma} +
c^\dagger_{j\sigma}c_{i\sigma}\right)\right.\\
\left. + \tilde{J} \sum_{\langle ij \rangle}
\mathbf{S}_i\cdot\mathbf{S}_j - \tilde{V} \sum_{\langle ij
\rangle} n_i n_j - \tilde{\mu} \sum_i n_i \right] P\,,
\end{multline}
where $c^\dagger_{i\sigma}$ and $c_{i\sigma}$ are creation and
annihilation operators, obeying anticommutation rules, for an
electron with spin $\sigma =\: \uparrow$ or $\downarrow$ at lattice
site $i$, $n_{i\sigma} = c^\dagger_{i\sigma}c_{i\sigma}$, $n_i =
n_{i\uparrow} + n_{i\downarrow}$ are the number operators, and
$\mathbf{S}_i = \sum_{\sigma\sigma^\prime} c^\dagger_{i\sigma}
\mathbf{s}_{\sigma\sigma^\prime} c_{i\sigma^\prime}$ is the
single-site spin operator, with $\mathbf{s}$ the vector of Pauli
spin matrices.  The entire Hamiltonian is sandwiched between
projection operators $P = \prod_{i}
(1-n_{i\downarrow}n_{i\uparrow})$, which project out states with
doubly-occupied sites.  The interaction constants $\tilde{t}$,
$\tilde{J}$, $\tilde{V}$ describe the following physical features:
electron hopping ($\tilde{t}$), a nearest-neighbor antiferromagnetic
coupling ($\tilde{J} >0$), and a nearest-neighbor interaction
($\tilde{V}$).  The standard $tJ$ Hamiltonian is a special case of
Eq.~\eqref{eq:1} with $\tilde{V}/\tilde{J} = 1/4$.  For convenience,
we introduce dimensionless interaction constants $t, J, V, \mu$, and
rearrange the $\tilde{\mu}$ chemical potential term to group the
Hamiltonian into a single lattice summation:
\begin{equation}
\label{eq:2} \begin{split} -\beta H = & \sum_{\langle ij \rangle}
P \biggl[ -t \sum_{\sigma} \left(c^\dagger_{i\sigma}c_{j\sigma} +
c^\dagger_{j\sigma}c_{i\sigma}\right)\\ &  - J
\mathbf{S}_i\cdot\mathbf{S}_j + V n_i n_j + \mu ( n_i +n_j)\biggr]
P\\
\equiv &\sum_{\langle ij \rangle} \{-\beta H (i,j)\}\,.
\end{split}
\end{equation}
Here $\beta = 1/k_B T$, so that the interaction constants are
related by $t = \beta \tilde{t}$, $J = \beta \tilde{J}$, $V= \beta
\tilde{V}$, $\mu = \beta\tilde{\mu}/2d$.  The total Hamiltonian is
now written as a sum of pair Hamiltonians $-\beta H(i,j)$. The sum
over nearest-neighbor sites $(i,j)$ is taken so that the position of
site $j$ is $\mathbf{r}_j = \mathbf{r}_i + \mathbf{a}_k$, where
$\mathbf{a}_k$ is one of the $d$ lattice vectors.  Since changing
the sign of $t$ is equivalent to redefining the phase at every other
site in the system, we shall choose $t>0$ with no loss of
generality. The effective temperature variable will be $1/t = k_B T
$, where we have taken $\tilde t = 1$ as the unit of energy.

In order to study the superfluid weight, we introduce periodic
boundary conditions, by considering the system as a ring in each
axis direction threaded by a magnetic flux.  We choose the vector
potential $\mathbf{A}$ associated with the flux to have components
$A/\sqrt{d}$ along each axis, so that the pair Hamiltonian becomes
\begin{multline}\label{eq:3}
-\beta H(i,j) = P \biggl[ -t \sum_{\sigma} \left(e^{i\phi}
c^\dagger_{i\sigma}c_{j\sigma} + e^{-i\phi}
c^\dagger_{j\sigma}c_{i\sigma}\right)\\ - J
\mathbf{S}_i\cdot\mathbf{S}_j + V n_i n_j + \mu ( n_i +n_j)\biggr]
P\,,
\end{multline}
where $\phi= a A/\sqrt{d}$ and $a$ is the lattice spacing.  For
simplicity, we have adopted units so that $\hbar = c = e = 1$.  In
the presence of the applied phase twist $\phi$, the superfluid
weight $n_s/m^\ast$ is related to the curvature of the total free
energy $F$ near $\phi=0$,\cite{Fisher, Scalapino}
\begin{equation}\label{eq:4}
\frac{n_s}{m^\ast} = \frac{1}{Na^2} \lim_{A \to 0} \frac{\partial^2
F}{\partial A^2} = \frac{1}{Nd} \lim_{\phi \to 0} \frac{\partial^2
F}{\partial \phi^2}\,,
\end{equation}
where $N \to \infty$ is the total number of lattice sites.  In
Sec.IIIE we shall show how this quantity can be calculated from the
renormalization-group transformation developed below.

\section{Renormalization-Group Transformation}

\subsection{Recursion Relations}

The position-space renormalization-group method used here starts
with an approximate decimation in $d=1$, which is then generalized
to higher dimensions by the Migdal-Kadanoff
procedure~\cite{FalicovBerkerT, FalicovBerker}.  In $d=1$, the
Hamiltonian of Eq.~\eqref{eq:2} takes the form:
\begin{equation}\label{eq:14}
-\beta H = \sum_i \left\{-\beta H(i,i+1) \right\}\,,
\end{equation}
where $i=1,2,3,\ldots$.  The decimation consists of finding a
thermodynamically equivalent system, described by the Hamiltonian
$-\beta^\prime H^\prime$, which depends only on the states of the
odd-numbered sites.  Since the quantum operators in the
Hamiltonian do not commute, an exact decimation even in one
dimension is not possible.  We can carry out an approximate
decimation as follows~\cite{SuzTak,TakSuz}:
\begin{equation}
\label{eq:15} \begin{split} \text{Tr}_{\text{even}} e^{-\beta H}
=&\text{Tr}_{\text{
even}}e^{\sum_{i}\left\{ -\beta H(i,i+1)\right\} }\\
=&\text{Tr}_{\text{even}} e^{\sum_{i}^{\text{
even}}\left\{ -\beta H(i-1,i)-\beta H(i,i+1) \right\} }\\
\simeq& \prod_{i}^{\text{even}}\text{Tr}_{i}e^{\left\{
-\beta H(i-1,i)-\beta H(i,i+1)\right\} }\\
=&\prod_{i}^{\text{
even}}e^{-\beta ^{\prime }H^{\prime }(i-1,i+1)}\\
\simeq& e^{\sum_{i}^{\text{even}}\left\{ -\beta ^{\prime
}H^{\prime }(i-1,i+1)\right\} } =e^{-\beta ^{\prime }H^{\prime }}.
\end{split}
\end{equation}
Here $-\beta^\prime H^\prime$ is the Hamiltonian for the
renormalized system, and $\text{Tr}_{\text{even}}$ is a trace over
the degrees of freedom at all even-numbered sites.  In the two
approximate steps, marked by $\simeq$ in Eq.~\eqref{eq:15}, we
ignore the non-commutation of operators separated beyond three
consecutive sites of the unrenormalized system (conversely, this
means that anticommutation rules are taken into account within three
consecutive sites at all successive length scales, as the
renormalization-group procedure is repeated).  These two steps
involve the same approximation but in opposite directions, which
gives some mutual compensation. Earlier studies of quantum spin
systems have shown the success of this approximation at predicting
finite-temperature behavior.\cite{SuzTak,TakSuz}

The renormalization-group mapping can be extracted from the third
and fourth lines of Eq.(\ref{eq:15}):
\begin{equation}
e^{-\beta ^{\prime }H^{\prime }(i,k)}=\mbox{Tr}_{j}e^{-\beta
H(i,j)-\beta H(j,k)}, \label{eq:16}
\end{equation}
where $i,j,k$ are three consecutive sites of the unrenormalized
system. The operators $-\beta ^{\prime }H^{\prime }(i,k)$ and
$-\beta H(i,j)-\beta H(j,k)$ act on the space of two-site and
three-site states respectively, so that, in terms of matrix
elements,
\begin{multline}
\langle u_{i}v_{k}|e^{-\beta ^{\prime }H^{\prime }(i,k)}|\bar{u}_{i}^{{}}%
\bar{v}_{k}^{{}}\rangle = \label{eq:17}\\
\sum_{w_{j}}\langle u_{i}\,w_{j}\,v_{k}|e^{-\beta H(i,j)-\beta
H(j,k)}|\bar{u}_{i}\,w_{j}\,\bar{v}_{k}^{{}}\rangle \:,
\end{multline}
where $u_{i},w_{j},v_{k},\bar{u}_{i},\bar{v}_{k}^{{}}$ are
single-site state variables. Eq.(\ref{eq:17}) is the contraction
of a $27\times 27$ matrix on the right into a $9\times 9$ matrix
on the left.  We block-diagonalize the left and right sides of
Eq.(\ref{eq:17}) by choosing basis states which are the
eigenstates of total particle number, total spin magnitude, total
spin $z$-component, and parity. We denote the set of 9 two-site
eigenstates by $\{|\phi _{p}\rangle \}$ and the set of 27
three-site eigenstates by $\{|\psi _{q}\rangle \}$, and list them
in Tables I and II.  Eq.(\ref{eq:17}) is rewritten as
\begin{multline}
\langle \phi _{p}|e^{-\beta ^{\prime }H^{\prime }(i,k)}|\phi _{\bar{p}%
}\rangle = \label{eq:18}\\
\sum_{\substack{u,v,\bar{u},\\ \bar{v},w}}
\sum_{\substack{q,\bar{q}}} \langle\phi _p|u_iv_k\rangle \langle
u_iw_jv_k|\psi_q\rangle \langle \psi _q|e^{-\beta H(i,j)-\beta
H(j,k)}|\psi _{\bar{q}}\rangle\cdot \\
\langle \psi_{\bar{q}}|\bar{u}_iw_j\bar{v}_k\rangle \langle
\bar{u}_i\bar{v}_k|\phi _{\bar{p}}\rangle\:.
\end{multline}

Eq.~\eqref{eq:18} yields six independent elements for the matrix
$\langle \phi _{p}|e^{-\beta ^{\prime }H^{\prime
}(i,k)}|\phi_{\bar{p}}\rangle$, which we label $\gamma_p$ as
follows:
\begin{equation}\label{eq:19}
\begin{split}
\gamma_p &\equiv \langle \phi _{p}|e^{-\beta ^{\prime }H^{\prime
}(i,k)}|\phi_{p}\rangle \quad \text{for}\: p = 1,2,4,6,7,\\
\gamma_0 &\equiv \langle \phi _{2}|e^{-\beta ^{\prime }H^{\prime
}(i,k)}|\phi_{4}\rangle\,.
\end{split}
\end{equation}
To calculate the $\gamma_p$, we determine the matrix elements of
$-\beta H(i,j) -\beta H(j,k)$ in the three-site basis $\{\psi_q\}$,
as listed in Table III, and exponentiate the matrix blocks to find
the elements $\langle \psi _q|e^{-\beta H(i,j)-\beta H(j,k)}|\psi
_{\bar{q}}\rangle$ which enter on the right-hand side of
Eq.~\eqref{eq:18}.  In this way the $\gamma_p$ are functions of the
interaction constants in the unrenormalized Hamiltonian, $\gamma_p =
\gamma_p (t,\phi,J,V,\mu)$.

\begin{table}[h]
\begin{tabular}{|c|c|c|c|c|}
 \hline
  $n$ & $p$ & $s$ & $m_s$ & Two-site basis states\\
  \hline
  $0$ & $+$ & $0$ & $0$ &$|\phi_{1}\rangle=|\circ\circ\rangle$ \\
  \hline
  $1$ & $+$ & $1/2$ & $1/2$ &$|\phi_{2}\rangle=\frac{1}{\sqrt{2}}\{|\uparrow
  \circ\rangle+|\circ\uparrow\rangle\}$\\ \hline
  $1$ & $-$ & $1/2$ & $1/2$ &$|\phi_{4}\rangle=\frac{1}{\sqrt{2}}\{|\uparrow
  \circ\rangle-|\circ\uparrow\rangle\}$\\ \hline
  $2$ & $-$ & $0$ & $0$ & $|\phi_{6}\rangle=\frac{1}{\sqrt{2}}\{|\uparrow\downarrow\rangle
  -|\downarrow\uparrow\rangle\}$\\ \hline
  $2$ & $+$ & $1$ & $1$ &
  $|\phi_{7}\rangle=|\uparrow\uparrow\rangle$\\ \hline
  $2$ & $+$ & $1$ & $0$ &
  $|\phi_{9}\rangle=\frac{1}{\sqrt{2}}\{|\uparrow\downarrow\rangle+|\downarrow
  \uparrow\rangle\}$\\ \hline
\end{tabular}
\caption{The two-site basis states, with the corresponding
particle number ($n$), parity ($p$), total spin ($s$), and total
spin $z$-component ($m_s$) quantum numbers.  The states
$|\phi_{3}\rangle$, $|\phi_{5}\rangle$, and $|\phi_{8}\rangle$,
are obtained by spin reversal from $|\phi_{2}\rangle$,
$|\phi_{4}\rangle$, and $|\phi_{7}\rangle$, respectively.}
\end{table}

\begin{table}
\begin{tabular}{|c|c|c|c|c|}
 \hline
  $n$ & $p$ & $s$ & $m_s$ & Three-site basis states\\
  \hline
  $0$ & $+$ & $0$ & $0$ &$|\psi_{1}\rangle=|\circ\circ\,\circ\rangle$ \\
  \hline
  $1$ & $+$ & $1/2$ & $1/2$ &$|\psi_{2}\rangle=|\circ
  \uparrow
  \circ\rangle,\: |\psi_{3}\rangle=\frac{1}{\sqrt{2}}\{|\uparrow
  \circ\,\circ\rangle+|\circ\,\circ\uparrow\rangle\}$\\ \hline
  $1$ & $-$ & $1/2$ & $1/2$ &$|\psi_{6}\rangle=\frac{1}{\sqrt{2}}\{|\uparrow
  \circ\,\circ\rangle-|\circ\,\circ\uparrow\rangle\}$\\ \hline
$2$ & $+$ & $0$ & $0$ &
  $|\psi_{8}\rangle=\frac{1}{2}\{|\uparrow\downarrow\circ\rangle-
  |\downarrow\uparrow\circ\rangle-|\circ\uparrow\downarrow\rangle+
  |\circ\downarrow\uparrow\rangle\}$\\ \hline
   $2$ & $-$ & $0$ & $0$ &
  $|\psi_{9}\rangle=\frac{1}{2}\{|\uparrow\downarrow\circ\rangle-
  |\downarrow\uparrow\circ\rangle+|\circ\uparrow\downarrow\rangle-
  |\circ\downarrow\uparrow\rangle\},$\\
  &&&&$|\psi_{10}\rangle=\frac{1}{\sqrt{2}}\{|\uparrow\circ\downarrow\rangle-|\downarrow\circ\uparrow
  \rangle\}$\\\hline
  $2$ & $+$ & $1$ & $1$ &
  $|\psi_{11}\rangle=|\uparrow\circ\uparrow\rangle,\:
  |\psi_{12}\rangle=\frac{1}{\sqrt{2}}\{|\uparrow\uparrow\circ\rangle+|\circ\uparrow\uparrow
  \rangle\}$\\ \hline
  $2$ & $+$ & $1$ & $0$ &
  $|\psi_{13}\rangle=\frac{1}{2}\{|\uparrow\downarrow\circ\rangle+
  |\downarrow\uparrow\circ\rangle+|\circ\uparrow\downarrow\rangle+
  |\circ\downarrow\uparrow\rangle\},$\\
  &&&& $|\psi_{14}\rangle=\frac{1}{\sqrt{2}}
  \{|\uparrow\circ\downarrow\rangle+|\downarrow\circ\uparrow
  \rangle\}$\\ \hline
  $2$ & $-$ & $1$ & $1$ &
  $|\psi_{17}\rangle=\frac{1}{\sqrt{2}}\{|\uparrow\uparrow\circ\rangle-|\circ\uparrow\uparrow
  \rangle\}$\\ \hline
  $2$ & $-$ & $1$ & $0$ &
  $|\psi_{18}\rangle=\frac{1}{2}\{|\uparrow\downarrow\circ\rangle+
  |\downarrow\uparrow\circ\rangle-|\circ\uparrow\downarrow\rangle-
  |\circ\downarrow\uparrow\rangle\}$\\ \hline
  $3$ & $+$ & $1/2$ & $1/2$ &
  $|\psi_{20}\rangle=\frac{1}{\sqrt{6}}\{2|\uparrow\downarrow\uparrow\rangle-|\uparrow\uparrow
  \downarrow\rangle-|\downarrow\uparrow\uparrow\rangle\}$\\
  \hline
  $3$ & $-$ & $1/2$ & $1/2$ &
  $|\psi_{22}\rangle=\frac{1}{\sqrt{2}}\{|\uparrow\uparrow
  \downarrow\rangle-|\downarrow\uparrow\uparrow\rangle\}$\\
  \hline
  $3$ & $+$ & $3/2$ & $3/2$ &
  $|\psi_{24}\rangle=|\uparrow\uparrow\uparrow\rangle$ \\ \hline
  $3$ & $+$ & $3/2$ & $1/2$ &
  $|\psi_{25}\rangle=\frac{1}{\sqrt{3}}\{|\uparrow\downarrow\uparrow\rangle+|\uparrow\uparrow
  \downarrow\rangle+|\downarrow\uparrow\uparrow\rangle\}$ \\ \hline
\end{tabular}
\caption{The three-site basis states, with the corresponding
particle number ($n$), parity ($p$), total spin ($s$), and total
spin $z$-component ($m_s$) quantum numbers. The states
$|\phi_{4-5}\rangle$, $|\phi_{7}\rangle$, $|\phi_{15-16}\rangle$,
$|\phi_{19}\rangle$, $|\phi_{21}\rangle$, $|\phi_{23}\rangle$,
$|\phi_{26-27}\rangle$, are obtained by spin reversal from
$|\phi_{2-3}\rangle$, $|\phi_{6}\rangle$, $|\phi_{11-12}\rangle$,
$|\phi_{17}\rangle$, $|\phi_{20}\rangle$, $|\phi_{22}\rangle$,
$|\phi_{24-25}\rangle$, respectively.}
\end{table}

\begingroup
\squeezetable
\begin{table}
\begin{gather*}
\begin{array}{|c||c|}\hline
 & \psi_{1}\\
\hhline{|=#=|} \psi_{1} & 0\\ \hline
\end{array}\\
\begin{array}{|c||c|c|c|}\hline
 & \psi_{2} & \psi_{3} & \psi_{6}\\
\hhline{|=#=|=|=|} \psi_{2} & 2\mu & -\sqrt{2}t \cos\phi & i \sqrt{2} t \sin\phi\\
\hline \psi_{3} & -\sqrt{2}t \cos\phi & \mu & i \Delta_1\\
\hline \psi_{6} & -i \sqrt{2} t \sin\phi & -i \Delta_1 & \mu\\
\hline
\end{array}\\
\begin{array}{|c||c|c|c|}\hline
 & \psi_{8} & \psi_{9} & \psi_{10}\\
\hhline{|=#=|=|=|} \psi_{8} & -\frac{3}{4}J + V + 3\mu  & -i
\Delta_3 &  -i \sqrt{2} t\sin \phi\\
\hline \psi_{9} & i \Delta_3 & -\frac{3}{4}J + V + 3\mu  & -\sqrt{2} t\cos \phi\\
\hline  \psi_{10} & i \sqrt{2} t\sin \phi &
   -\sqrt{2} t\cos \phi  & 2\mu \\
\hline
\end{array}\\
\begin{array}{|c||c|c|c|}\hline
 & \psi_{11} & \psi_{12} & \psi_{17}\\
\hhline{|=#=|=|=|} \psi_{11} & 2\mu &
-\sqrt{2} t \cos\phi &  i \sqrt{2} t\sin \phi\\
\hline \psi_{12} & -\sqrt{2}t\cos\phi & \frac{1}{4}J + V + 3\mu & i \Delta_2\\
\hline  \psi_{17} & -i \sqrt{2} t\sin \phi &
   -i\Delta_2  & \frac{1}{4}J + V + 3\mu  \\
\hline
\end{array}\\
\begin{array}{|c||c|c|c|}\hline
 & \psi_{13} & \psi_{14} & \psi_{18}\\
\hhline{|=#=|=|=|} \psi_{13} & \frac{1}{4}J + V + 3\mu &
-\sqrt{2} t \cos\phi &  i \Delta_4\\
\hline \psi_{14} & -\sqrt{2}t\cos\phi & 2\mu & i \sqrt{2}t\sin\phi\\
\hline  \psi_{18} & -i \Delta_4 &
   -i\sqrt{2} t\sin \phi & \frac{1}{4}J + V + 3\mu  \\
\hline
\end{array}\\
\begin{array}{|c||c|}\hline
 & \psi_{20}\\
\hhline{|=#=|} \psi_{20} & -J + 2V + 4\mu \\ \hline
\end{array}\quad
\begin{array}{|c||c|}\hline
 & \psi_{22}\\
\hhline{|=#=|} \psi_{22} & 2V + 4\mu \\ \hline
\end{array}\\
\begin{array}{|c||c|}\hline
 & \psi_{24}\\
\hhline{|=#=|} \psi_{24} & \frac{1}{2}J +2V +4\mu \\ \hline
\end{array}\quad
\begin{array}{|c||c|}\hline
 & \psi_{25}\\
\hhline{|=#=|} \psi_{25} & \frac{1}{2}J +2V +4\mu \\ \hline
\end{array}
\end{gather*}
\caption{Diagonal matrix blocks of the unrenormalized three-site
Hamiltonian $-\beta H(i,j)-\beta H(j,k)$.  The Hamiltonian being
invariant under spin-reversal, the spin-flipped matrix elements
are not shown.  The additive constant contribution $2G$, occurring
at the diagonal terms, is also not shown.  The additional
$\Delta_i$ terms, which are not part of the original three-site
Hamiltonian, are explained in Sec.IIIC.}
\end{table}
\endgroup
\begingroup
\squeezetable
\begin{table}
\begin{gather*}
\begin{array}{|c||c|c|c|c|c|c|} \hline & \parbox{0.08in}{$\phi_{1}$} &\phi_{2} &\phi_{4} & \phi_{6} &
\phi_{7} &\phi_{9}\\
\hhline{|=#=|=|=|=|=|=|} \parbox{0.1in}{$\phi_{1}$} & \parbox{0.08in}{$G^\prime$} & \multicolumn{5}{c|}{}\\
\cline{1-4} \parbox{0.1in}{$\phi_{2}$} && \parbox{0.5in}{\centering $-t^\prime\cos\phi^\prime+\mu^\prime+G^\prime$} & i t^\prime \sin\phi^\prime & \multicolumn{3}{c|}{0}\\
\cline{1-1}\cline{3-4} \parbox{0.1in}{$\phi_{4}$} &
\multicolumn{1}{c|}{} & -it^\prime \sin\phi^\prime
&\parbox{0.5in}{\centering $t^\prime\cos\phi^\prime + \mu^\prime+G^\prime$} &\multicolumn{3}{c|}{}\\
\cline{1-1}\cline{3-5} \parbox{0.1in}{$\phi_{6}$} &
\multicolumn{3}{c|}{} &
\multicolumn{1}{c|}{\parbox{0.575in}{\centering
$-\frac{3}{4}J^\prime +V_2^\prime+2\mu^\prime+G^\prime$}} &
\multicolumn{2}{c|}{}\\
\cline{1-1}\cline{5-6} \parbox{0.1in}{$\phi_{7}$} &
\multicolumn{4}{c|}{0}
&\multicolumn{1}{c|}{\parbox{0.485in}{\centering $\frac{1}{4}J^\prime+V_2^\prime+2\mu^\prime+G^\prime$}}&\\
\cline{1-1}\cline{6-7} \parbox{0.1in}{$\phi_{9}$} &
\multicolumn{5}{c|}{}
& \parbox{0.485in}{\centering $\frac{1}{4}J^\prime+V_2^\prime+2\mu^\prime+G^\prime$}\\
\hline
\end{array}
\end{gather*}
\caption{Block-diagonal matrix of the renormalized two-site
Hamiltonian $-\beta^\prime H^\prime(i,k)$.  The Hamiltonian being
invariant under spin-reversal, the spin-flipped matrix elements
are not shown.}
\end{table}
\endgroup

Since the matrix $\langle \phi _{p}|e^{-\beta ^{\prime }H^{\prime
}(i,k)}|\phi_{\bar{p}}\rangle$ is determined by six independent
elements $\gamma_p$, the renormalized pair Hamiltonian $-\beta
^{\prime }H^{\prime }(i,k)$ involves six interaction constants,
namely those of the original types of interactions and an additive
constant:
\begin{multline}\label{eq:20}
-\beta^\prime H^\prime(i,k) = P \biggl[ -t^\prime \sum_{\sigma}
\left(e^{i\phi^\prime} c^\dagger_{i\sigma}c_{j\sigma} +
e^{-i\phi^\prime} c^\dagger_{j\sigma}c_{i\sigma}\right)\\ -
J^\prime \mathbf{S}_i\cdot\mathbf{S}_j + V^\prime n_i n_j +
\mu^\prime ( n_i +n_j) + G^\prime\biggr] P\,,
\end{multline}
The matrix elements of $-\beta ^{\prime }H^{\prime }(i,k)$ in the
$\{\phi_p\}$ basis are shown in Table IV. Exponentiating this
matrix, we can solve for the renormalized interaction constants
$(t^\prime,\phi^\prime,J^\prime,V^\prime,\mu^\prime,G^\prime)$ in
terms of the $\gamma_p$:
\begin{gather}
t^\prime = \text{sign}\,(\gamma_4-\gamma_2)\cosh^{-1}
\left(\frac{\gamma_2+\gamma_4}{2e^{v}}\right),\nonumber\\[5pt]
\phi^\prime = \tan^{-1} \left(\frac{2\,
\text{Im}\,\gamma_0}{\gamma_4-\gamma_2}\right),\qquad  J^\prime =
\ln\frac{\gamma_7}{\gamma_6},\nonumber\\[5pt]
V^\prime = \frac{1}{4}\left\{\ln(\gamma_1^4 \gamma_6
\gamma_7^3)-8v\right\},\qquad  \mu^\prime = v - \ln
\gamma_1,\nonumber\\[5pt]
G^\prime=\ln\gamma_1,\label{eq:21}
\end{gather}
where
\begin{gather*}
v = \frac{1}{2}\ln\left(\gamma_2\gamma_{4}-\gamma_0^\ast
\gamma_0\right)\,.
\end{gather*}

The approximate $d=1$ decimation contained in
Eqs.~\eqref{eq:18}-\eqref{eq:21} can be expressed as a mapping of a
Hamiltonian with interaction constants $\mathbf{K} = \{
G,t,J,V,\mu,\phi\}$ onto another Hamiltonian with interactions
constants
\begin{equation}\label{eq:22}
\mathbf{K}^\prime = \mathbf{R}(\mathbf{K})\,.
\end{equation}
The Migdal-Kadanoff procedure~\cite{Migdal,Kadanoff} is used to
construct the renormalization-group transformation for $d>1$.  We
ignore a subset of the nearest-neighbor interactions in the
$d$-dimensional hypercubic lattice, leaving behind a new
$d$-dimensional hypercubic lattice where each point is connected to
its neighbor by two consecutive nearest-neighbor segments of the
original lattice. We apply the decimation described above to the
middle site between the two consecutive segments, giving the
renormalized nearest-neighbor couplings for the points in the new
lattice.  We compensate for the interactions that are ignored in the
original lattice by multiplying by a factor of $b^{d-1}$ the
interactions after the decimation, $b=2$ being the length rescaling
factor. Thus, the renormalization-group transformation of
Eq.~\eqref{eq:22} generalizes, for $d>1$, to
\begin{equation} \label{eq:13}
\mathbf{K^\prime}=b^{d-1} \mathbf{R}(\mathbf{K}).
\end{equation}

\subsection{Renormalization-Group Transformation
in the Presence of Magnetic Flux}

In order to correctly model the response of the system to an applied
magnetic flux, the renormalization-group approximation described in
the last two sections needs to be extended.  To see this, we first
review the formalism for calculating thermodynamic densities from
the renormalization-group flows.\cite{BerkerOstlundPutnam} Conjugate
to each interaction $K_\alpha$ of $\mathbf{K} = \{ K_\alpha\}$,
there is a density $M_\alpha$ (e.g., kinetic energy, electron
density),
\begin{equation}\label{eq:23a}
M_\alpha = \frac{1}{Nd} \frac{\partial \ln Z}{\partial
K_\alpha}\,,
\end{equation}
where $Z(\mathbf{K})$ is the partition function.  We can relate the
densities at the two consecutive points along a
renormalization-group trajectory by
\begin{equation}\label{eq:23b}
M_\alpha = b^{-d} M^\prime_\beta T_{\beta \alpha}\,, \quad
\text{where} \qquad T_{\beta\alpha} \equiv \frac{\partial
K^\prime_\beta}{\partial K_\alpha}\,,
\end{equation}
with summation over repeated indices implied.  At a fixed point of
the renormalization-group transformation, corresponding to a phase
transition or a phase sink, the densities $M_\alpha =
M_\alpha^\prime \equiv M_\alpha^\ast$ are the left eigenvector with
eigenvalue $b^d$ of the recursion matrix $\mathbf{T}$ evaluated at
the fixed point. The densities at the starting point of the
trajectory (the actual physical system) are computed by iterating
Eq.~\eqref{eq:23b} until a fixed point is effectively reached.  If
$\mathbf{T}^{(k)}$ is the recursion matrix of the $(k)$th
renormalization-group iteration, then for large $k$, we can express
the densities of the actual system $\mathbf{M}$ as
\begin{equation}\label{eq:25}
\mathbf{M} \simeq b^{-kd} \mathbf{M}^\ast\ \cdot [\mathbf{T}^{(k)}]
\cdot [\mathbf{T}^{(k-1)}] \cdot \cdots \cdot [\mathbf{T}^{(1)}].
\end{equation}

The renormalization-group transformation incorporated in
Eqs.~\eqref{eq:18}-\eqref{eq:13} gives
\begin{gather}
\frac{\partial t^\prime}{\partial \phi} = \frac{\partial
J^\prime}{\partial \phi} = \frac{\partial V^\prime}{\partial \phi}
=\frac{\partial G^\prime}{\partial \phi}= 0\,,\nonumber\\
\frac{\partial \phi^\prime}{\partial t} = \frac{\partial
\phi^\prime}{\partial J} = \frac{\partial \phi^\prime}{\partial V}
=\frac{\partial \phi^\prime}{\partial \mu}= 0\,, \quad
\frac{\partial \phi^\prime}{\partial \phi} = 2\,,\label{eq:28}
\end{gather}
for all $\phi$.  The $6\times 6$ recursion matrix $\mathbf{T}$
will then have the form
\begin{equation}\label{eq:29}
\mathbf{T} = \left(\begin{array}{cccc|c} b^d & \frac{\partial
G^\prime}{\partial t} &
\cdots & \frac{\partial G^\prime}{\partial \mu}  &  \\
0  & \frac{\partial t^\prime}{\partial t} & \cdots &
\frac{\partial t^\prime}{\partial \mu}  & 0 \\
\vdots & \vdots & \ddots & \vdots &  \\
0 & \frac{\partial \mu^\prime}{\partial t} & \cdots &
\frac{\partial \mu^\prime}{\partial \mu}  &  \\
\hline & 0 & & & 2\\
\end{array}\right)
\end{equation}
at every step in the flow.  This leads to
\begin{equation}\label{eq:30}
M^\ast_6 = 0\ \quad \text{and} \quad \frac{\partial}{\partial \phi}
\ln Z = M_6 =0\,,
\end{equation}
for all points of the phase diagram.  This superfluid weight of zero
for all temperatures and electronic densities is clearly due to the
oversimplification in our initial approximation.

The source of the problem is the three-site cluster approximation
used in deriving the recursion relations.  In modifying the original
approximation scheme, we seek to incorporate the effect of the
non-commutations extending beyond the three-site cluster. Turning to
the matrix elements of $-\beta H(i,j) -\beta H(j,k)$ listed in Table
III, we note the terms $\Delta_i$, $i=1,\ldots,4$. Using the
original Hamiltonian of Eq.~\eqref{eq:3} restricted to the
three-cluster, the matrix elements involving these terms are all
zero.  However, non-commutativity extending beyond the three-cluster
makes, as we see below, these matrix elements non-zero.

We can estimate the magnitude of the matrix elements $\Delta_i$ by
considering a five-site cluster, described by Hamiltonian $-\beta
H(h,i) -\beta H(i,j) -\beta H(j,k) -\beta H(k,l)$, where
$(h,i,j,k,l)$ are consecutive sites.  In the spirit of
Eq.~\eqref{eq:17}, we generate effective couplings for the
three-cluster by tracing over the degrees of freedom at the outside
sites in the five-cluster,
\begin{multline}
\langle u_{i}\,v_{j}\, w_{k}|e^{-\tilde{\beta} \tilde{H}(i,j,k)}|\bar{u}_{i}\,
 \bar{v}_{j}\,\bar{w}_{k} \rangle = \label{eq:32}\\
\sum_{t_{h},x_{l}}\langle t_{h}\,
u_{i}\,v_{j}\,w_{k}\,x_{l}|e^{-\beta H(h,i) -\beta H(i,j)-\beta
H(j,k)-\beta
H(k,l)}\\
\cdot|t_h\,\bar{u}_{i}\,\bar{v}_{j}\,\bar{w}_{k}\,x_l\rangle \:,
\end{multline}
where the subscripted variables refer to single-site states. From
the above equation, we can extract the matrix elements of an
effective three-cluster Hamiltonian $-\tilde{\beta}
\tilde{H}(i,j,k)$.  Eq.~\eqref{eq:32} is the contraction of a $243
\times 243$ matrix on the right-hand side into a $27 \times 27$
matrix on the left.  We simplify our task by using the $\{\psi_p\}$
basis on the left, and choosing an appropriate five-site basis to
block-diagonalize the $243 \times 243$ right-hand matrix.

Since $-\tilde{\beta} \tilde{H}(i,j,k)$ is derived from the
decimation of a five-cluster, it will have a more general form than
$-\beta H(i,j) -\beta H(j,k)$, and approximately reflect the effect
of the three-cluster non-commutations with the external sites.
However our approximation scheme must also satisfy an important
constraint: the $\phi \to 0$ limit should yield the same
renormalization-group transformation used in earlier studies of the
$tJ$ model~\cite{FalicovBerkerT,FalicovBerker}.  To achieve this, we
modify only a subset of the matrix elements of $-\beta H(i,j) -\beta
H(j,k)$, namely those which are zero in the original scheme when
$\phi \ne 0$, but whose corresponding elements in $-\tilde{\beta}
\tilde{H}(i,j,k)$ are non-zero:
\begin{equation}\label{eq:33}
\begin{split}
\Delta_1 &= \text{sign}(\phi t)|\langle \psi_3 |
\tilde{\beta}\tilde{H}(i,j,k) |\psi_6\rangle|,\\
\Delta_2 &= \text{sign}(\phi t)|\langle \psi_{12} |
\tilde{\beta}\tilde{H}(i,j,k) |\psi_{17}\rangle|,\\
\Delta_3 &= \text{sign}(\phi t)|\langle \psi_{8} |
\tilde{\beta}\tilde{H}(i,j,k) |\psi_{9}\rangle|,\\
\Delta_4 &= \text{sign}(\phi t)|\langle \psi_{13} |
\tilde{\beta}\tilde{H}(i,j,k) |\psi_{18}\rangle|.
\end{split}
\end{equation}
The $\text{sign}(\phi t)$ prefactors guarantee that couplings
between the same types of three-cluster states have the same sign.
For example, $|\psi_2\rangle$ and $|\psi_3\rangle$ share the same
$n$, $p$, $s$, and $m_s$ quantum numbers, as can be seen from Table
II. A nonzero $\phi$ couples $|\psi_2\rangle$ to $|\psi_6\rangle$, a
state with the same $n$, $s$, and $m_s$, but opposite parity. From
the second block in Table III, the associated matrix element is
$\langle \psi_2 |\cdots|\psi_6\rangle = i\sqrt{2}t\sin(\phi)$. The
$\Delta_1$ elements in that block have an analogous role, coupling
$|\psi_3\rangle$ to $|\psi_6\rangle$. The prefactor in the
$\Delta_1$ expression of Eq.~\eqref{eq:33} sets the sign of the
element $\langle \psi_3 |\cdots|\psi_6\rangle = i\Delta_1$ to equal
that of $\langle \psi_2 |\cdots|\psi_6\rangle$.  Since our
calculations are all done for small $\phi$, $\text{sign}(\sin \phi)
= \text{sign}(\phi)$. Similar reasoning applies to the prefactors of
the other $\Delta_i$ elements.

Through Eq.~\eqref{eq:32}, the $\Delta_i$ are functions of the
interactions strengths in the unrenormalized Hamiltonian, $\Delta_i
= \Delta_i(t,J,V,\mu,\phi)$.  They scale like $\phi$ for small
$\phi$, and duly vanish in the limit $\phi \to 0$.  As will be
explained in Sec.IIIE, finding the superfluid weight involves
calculating a thermodynamic density in the $\phi \to 0$ limit, so we
shall be working in the regime where the $\Delta_i$ are vanishingly
small.  The result of the extended calculation, taking into account
the quantum mechanical backflow into the three-cluster, is that
Eqs.~\eqref{eq:28} no longer hold, $\partial \ln Z /\partial \phi
\ne 0$ in general, and we obtain interesting nontrivial results for
$n_s/m^\ast$.

\subsection{Calculation of the Superfluid Weight}

The superfluid weight of Eq.~\eqref{eq:4} is expressed as a
derivative of the total free energy $F = F(n,T,\phi)$, where $n =
\langle n_i \rangle$ is the electron density. In terms of the
conjugate current
\begin{equation}\label{eq:34}
j(n,T,\phi) = \frac{1}{Nd}\left.\frac{\partial F}{\partial
\phi}\right|_{n,T}\,,
\end{equation}
Eq.~\eqref{eq:4} becomes
\begin{equation}\label{eq:35}
\frac{n_s}{m^\ast}(n,T) = \lim_{\phi \to 0} \left.\frac{\partial
j}{\partial \phi}\right|_{n,T}\,.
\end{equation}
In terms of the grand potential $\Omega(\mu,T,\phi) = -(1/\beta)\ln
Z$,
\begin{equation}\label{eq:36}
j(\mu,T,\phi) = \frac{1}{Nd} \left.\frac{\partial \Omega}{\partial
\phi}\right|_{\mu,T}\,,
\end{equation}
and
\begin{equation}\label{eq:40}
n(\mu,T,\phi) = -\frac{\beta}{2Nd} \left.\frac{\partial
\Omega}{\partial \mu}\right|_{T,\phi}\,.
\end{equation}
Relating the partial derivatives of $j$ with respect to $\phi$
through
\begin{equation}\label{eq:37}
\left.\frac{\partial j}{\partial \phi}\right|_{\mu,T} =
\left.\frac{\partial j}{\partial n}\right|_{\phi,T}
\left.\frac{\partial n}{\partial \phi}\right|_{\mu,T} +
\left.\frac{\partial j}{\partial \phi}\right|_{n,T}\,,
\end{equation}
and using the Maxwell relation $\left.\frac{\partial n}{\partial
\phi}\right|_{\mu,T} = -\frac{\beta}{2}\left.\frac{\partial
j}{\partial \mu}\right|_{\phi,T}$,
\begin{equation}\label{eq:38}
\left.\frac{\partial j}{\partial \phi}\right|_{\mu,T} =
-\frac{\beta}{2}\left.\frac{\partial j}{\partial
n}\right|_{\phi,T} \left.\frac{\partial j}{\partial
\mu}\right|_{\phi,T} + \left.\frac{\partial j}{\partial
\phi}\right|_{n,T}\,.
\end{equation}
The current $j$ is zero when $\phi=0$, so that the first term on
the right-hand side above is also zero in the limit $\phi \to 0$,
and we find that $\lim_{\phi\to 0} \left.\frac{\partial
j}{\partial \phi}\right|_{\mu,T} = \lim_{\phi\to 0}
\left.\frac{\partial j}{\partial \phi}\right|_{n,T}$.  Thus
Eq.~\eqref{eq:35} can be equivalently written as
\begin{multline}\label{eq:39}
\frac{n_s}{m^\ast}(\mu,T) = \lim_{\phi \to 0} \left.\frac{\partial
j}{\partial \phi}\right|_{\mu,T} = \frac{1}{Nd} \lim_{\phi \to 0}
\left.\frac{\partial^2 \Omega}{\partial \phi^2}\right|_{\mu,T}\\
=-\frac{1}{\beta Nd} \lim_{\phi \to 0} \left.\frac{\partial^2 \ln
Z}{\partial \phi^2}\right|_{\mu,T} \,.
\end{multline}
This is the form we shall use when calculating the superfluid
weights.

\section{Results}

\subsection{Global Phase Diagram for $d=3$}

Each sink, or completely stable fixed point of the
renormalization-group flows, corresponds to a thermodynamic phase,
and we find the global phase diagram by determining the basin of
attraction for every sink~\cite{BerkerWortis}.  Flows that start at
the boundaries between phases have their own fixed points,
distinguished from phase sinks by having at least one unstable
direction.  Analysis of these fixed points determines whether the
phase transition is first- or second-order.  As explained in
Sec.IIIC, the thermodynamic densities, which are the expectation
values of operators occurring in the Hamiltonian, can also be
calculated from the renormalization-group flows.  In particular, we
determine the single-site electron density $\langle n_i \rangle$.
For the coupling $J/t = 0.444$ and $\phi =0$, the phase diagram in
terms of $\langle n_i \rangle$ and temperature $1/t$ is shown in
Fig. 1 \cite{FalicovBerkerT,FalicovBerker}.

\begin{figure}
\centering
\includegraphics*[scale=1]{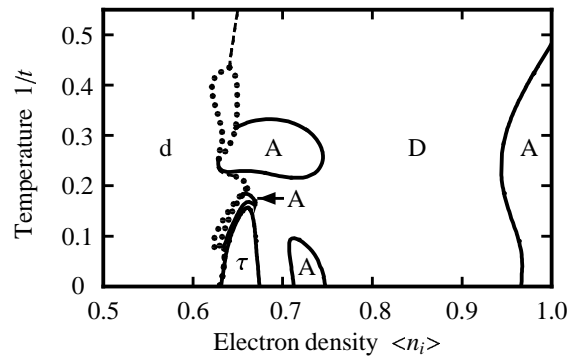}
\caption{Phase diagram for the $d=3$ $tJ$ model with $J/t = 0.444$,
$\phi = 0$, in temperature versus electron
density.\cite{FalicovBerker}  The antiferromagnetic (A), dense
disordered (D), dilute disordered (d), and $\tau$ phases are seen.
The second-order phase boundaries are drawn with full curves.  The
coexistence boundaries of first-order transitions are drawn with
dotted curves, with the unmarked areas inside corresponding to
coexistence regions of the two phases at either side.  The dashed
lines are not phase transitions, but disorder lines between the
dilute disordered and dense disordered phases.}
\end{figure}

\renewcommand{\arraystretch}{1.3}
\begin{table}[h]
\begin{tabular}{|c|c|c|c|c|}
 \hline
  Phase sink & \multicolumn{4}{c|}{Expectation values}\\
  \cline{2-5}
  & $\sum_\sigma \langle c^\dagger_{i\sigma}c_{j\sigma} +
c^\dagger_{j\sigma}c_{i\sigma}\rangle$  & $\langle n_i \rangle$ &
$\langle \mathbf{S}_i \cdot
\mathbf{S}_j \rangle$ & $\langle n_i n_j \rangle$\\
  \hline
  d & 0 & 0 & 0 & 0\\ \hline
  D & 0 & 1 & 0 & 1\\ \hline
  A & 0 & 1 & $\frac{1}{4}$ & 1 \\\hline
  $\tau$ & $-\frac{2}{3}$ & $\frac{2}{3}$ & $-\frac{1}{4}$ & $\frac{1}{3}$ \\\hline
\end{tabular}
\caption{Expectation values at the phase-sink fixed points.}
\end{table}

The nature of the various phases is epitomized by the thermodynamic
densities $\mathbf{M}^\ast$ calculated at each phase sink (Table V),
which underpin the calculation of densities throughout their
respective phases (Eq.~\eqref{eq:25}). We summarize the phase
properties below (for a more detailed discussion,
see~\cite{FalicovBerkerT,FalicovBerker}):

{\bf Dilute disordered phase (d):}  The electron density $\langle
n_i \rangle = 0$ at the sink and, as a result, the $\langle n_i
\rangle$ calculated inside this phase are low.

{\bf Dense disordered phase (D):}  The electron density $\langle n_i
\rangle = 1$ at the sink and, as a result, the $\langle n_i \rangle$
calculated inside this phase are close to 1.

{\bf Antiferromagnetic phase (A):} The electron density $\langle n_i
\rangle = 1$ at the sink, so that this phase is also densely filled.
The nearest-neighbor spin-spin correlation $\langle \mathbf{S}_i
\cdot \mathbf{S}_j\rangle = 1/4$ at the sink. Two spins that are
nearest neighbors at the sink are distant members of the same
sublattice in the original cubic lattice. The non-zero value of the
correlation function at the sink leads to $\langle \mathbf{S}_i
\cdot \mathbf{S}_j\rangle < 0$ for nearest-neighbor sites of the
original, unrenormalized system.

{\bf $\tau$ phase:}  This is a novel phase, characterized by
partial-filling at the phase sink, $\langle n_i\rangle  = 2/3$.  It
is the only phase where the electron hopping strength $t$ does not
renormalize to zero after repeated rescalings; instead, $t \to
\infty$ at the sink.  As a result, the expectation value of the
electron hopping operator at the sink is non-zero, $\sum_\sigma
\langle c^\dagger_{i\sigma}c_{j\sigma} +
c^\dagger_{j\sigma}c_{i\sigma}\rangle = -2/3$.  This property makes
it a possible $tJ$ model analogue to the superconducting phase in
high-$T_c$ materials.  The superfluid weight and thermodynamic
results discussed below certainly support this idea.

In the limit $\langle n_i \rangle = 1$, the system exhibits
antiferromagnetic order at low temperatures, as expected from the
spin-spin coupling in the Hamiltonian.  Upon hole doping, there is a
competition between the A and D phases, which respectively minimize
antiferromagnetic potential energy and hole kinetic energy.  Note
the extent of the A phase near $\langle n_i \rangle = 1$, which
persists only up to a small amount of hole doping $\delta = 1-
\langle n_i \rangle \lesssim 0.05$.  This feature is directly
reminiscent of the antiferromagnetic phase in certain high-$T_c$
materials, for example La$_{2-x}$Sr$_x$CuO$_4$~\cite{Imada}.  At
intermediate dopings $\delta \approx 0.32 - 0.37$, we have a
low-temperature $\tau$ phase, surrounded by islands of
antiferromagnetism. (When the hopping strength $t$ increases under
rescaling, this also lowers the free energy of antiferromagnetically
long-range ordered states, which leads to these islands of A in the
vicinity of the $\tau$ phase.\cite{FalicovBerker})  At hole dopings
$\delta \gtrsim 0.37$, there is a transition to a dilute disordered
phase, with a narrow region of first-order coexistence at lower
temperatures.

\subsection{Superfluid Weight and Kinetic Energy}

Using the method of calculating thermodynamic densities described in
Sec.IIIE, we determine $(1/Nd) \partial \ln Z/\partial \phi$ at
small $\phi$.  Taking the numerical derivative of this quantity at
$\phi=0$, we find $n_s/m^\ast$ through Eq.~\eqref{eq:39}.  The
superfluid weight is plotted as a function of electron density in
Fig. 2, along four different constant temperature cross-sections of
the phase diagram.  For comparison, we also show in the same figure
the calculated average kinetic energy per bond $\langle K \rangle$,
where $K = - \sum_\sigma \left( c^\dagger_{i\sigma} c_{j\sigma}+
c^\dagger_{j\sigma} c_{i\sigma}\right)$.  $\langle K \rangle$ and
the total weight of $\sigma_1(\omega,T)$, the real part of the
optical conductivity, are related by the sum rule~\cite{Tan},
\begin{equation}\label{eq:8}
\int_0^\infty d\omega\, \sigma_1(\omega,T) = \frac{\pi e^2}{2}
\langle K \rangle\,.
\end{equation}
In comparing the properties of the $tJ$ model to those of
high-$T_c$ materials, we keep in mind that the $tJ$ Hamiltonian
describes a one-band model, so cannot account for interband
transitions. For real materials, the full conductivity sum rule
has the form
\begin{equation}\label{eq:10}
\int_0^\infty d\omega\, \sigma_1(\omega,T) = \frac{\pi e^2
n}{2m}\,,
\end{equation}
where $n$ is the total density of electrons and $m$ is the free
electron mass.  The right-hand side of Eq.~\eqref{eq:10} is
independent of electron-electron interactions, in contrast to the
right-hand side of Eq.~\eqref{eq:8}, where $\langle K\rangle$ will
vary with the interaction strengths in the Hamiltonian.  The optical
conductivity of actual materials incorporates both transitions
within the conduction band and those to higher bands, while the $tJ$
model contains only the conduction band.  We can look at
Eq.~\eqref{eq:8} as a partial sum rule~\cite{Baeriswyl,Tan}, which
reflects the spectral weight of the free carriers in the conduction
band.

The experimental quantity we are interested in modeling is the
effective density of free carriers, which in actual materials is
calculated from the low-frequency spectral
weight~\cite{Orenstein},
\begin{equation}\label{eq:10b}
n_{\text{free}}(T) = \frac{2m}{\pi e^2} \int_0^{\omega_0}
d\omega\,\sigma_1(\omega,T)\,.
\end{equation}
For high $T_c$ materials, the cutoff frequency is typically chosen
around $\hbar \omega_0 \approx 1$ eV so as to include only
intraband transitions.  For comparison with the $tJ$ model, we
identify the right-hand side of Eq.~\eqref{eq:8} with

\begin{equation}\label{eq:10b}
\begin{split}
\frac{\pi e^2}{2} \langle K \rangle &= \frac{\pi e^2
n_{\text{free}}(T)}{2m}\,.
\end{split}
\end{equation}
The superfluid weight satisfies the inequality~\cite{Paramekanti}
\begin{equation}\label{eq:41}
\frac{n_s}{m^\ast} \le \langle K \rangle =
\frac{n_{\text{free}}}{m}\,,
\end{equation}
which is obeyed in our results in Fig.2.

\begin{figure} \centering

\includegraphics*[scale=0.95]{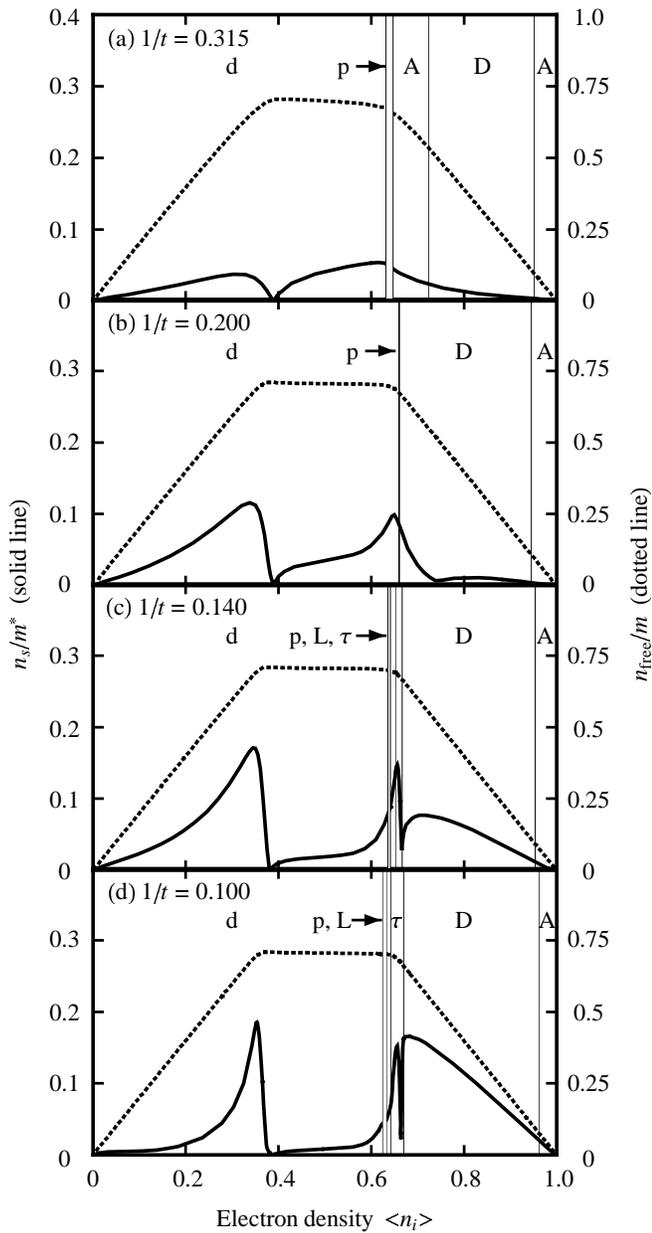}
\caption{The superfluid weight $n_s/m^\ast$ (solid line) and free
carrier density $n_{free}/m$ (dotted line) as a function of electron
density at four different values of temperature $1/t$. The
corresponding phases are indicated above the plots, and the location
of phase boundaries marked by thin vertical lines.  The symbol p
refers to a region of forbidden densities due to the discontinuity
at a first-order transition. The symbol L refers to a ``lamellar''
region where narrow slivers of the A and D phases alternate.}
\end{figure}

\begin{figure} \centering

\includegraphics*[scale=1.0]{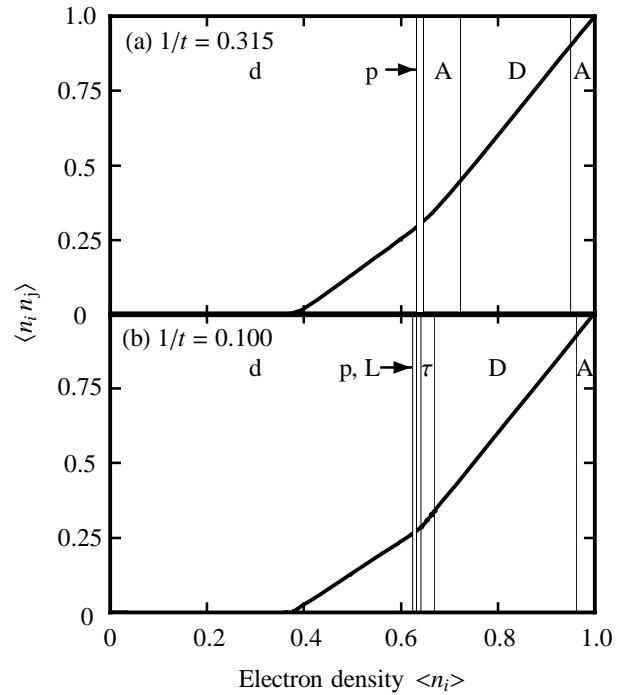}
\caption{The nearest-neighbor density-density correlation $\langle
n_i n_j \rangle$ as a function of electron density at two
different values of temperature $1/t$.  The corresponding phases
are indicated above the plots, and the location of phase
boundaries marked by thin vertical lines.  The symbol p refers to
a region of forbidden densities due to the discontinuity at a
first-order transition. The symbol L refers to a ``lamellar''
region where narrow slivers of the A and D phases alternate.}
\end{figure}

The superfluid weight graphs at the sampled temperatures show a
clear bipartite structure, with a peak at low $\langle n_i \rangle$,
and another peak at high $\langle n_i \rangle$ (which develops into
two closely spaced peaks at lower temperatures). In between these is
a region of low superfluid weight, with a minimum near $\langle
n_i\rangle \simeq 0.385$, approximately independent of temperature.
Looking at the nearest-neighbor density-density correlation $\langle
n_i n_j \rangle$ as shown in Fig. 3, we see that $\langle n_i
\rangle \simeq 0.385$ is also the electron density separating two
different regimes of the system: an extremely dilute regime, where
$\langle n_i n_j \rangle \simeq 0$, and a partially-to-densely
filled regime, where $\langle n_i n_j \rangle > 0$.  It is therefore
useful to discuss the superfluid weight and kinetic energy results
in terms of these two regimes.

\subsubsection{Extremely dilute regime, $\langle n_i\rangle \lesssim 0.385$}

The system in this regime is a dilute gas of electrons.  For low
$\langle n_i \rangle$, the kinetic energy per bond $\langle K
\rangle \simeq 2 \langle n_i \rangle$, which follows if the density
of free carriers is just the density of electrons, $n_\text{free} =
\langle n_i \rangle$, and the mass of the carriers $m = 1/2$. The
interaction terms in the $tJ$ Hamiltonian create an attractive
potential of strength $-\tilde{J}$ between electrons in
singlet-states on neighboring sites.  For a coupling $J/t = 0.444$,
this attraction is too weak to form two-body bound states, but since
we are in three dimensions, even a weak attractive potential is
sufficient for the formation of an electron superfluid at low
temperatures~\cite{Emery, Randeria}.  In fact, we see a peak in
$n_s/m^\ast$ develop around $\langle n_i \rangle \approx$ 0.3--0.35,
and this peak grows as the temperature is lowered from $1/t = 0.315$
to $0.1$.  For low $\langle n_i\rangle$, the superfluid weight
increases with electron density and $\langle K \rangle$.  The
location of the peak in $n_s/m^\ast$ is just before $\langle K
\rangle$ comes to its maximum and levels off. As the density of free
carriers saturates near $\langle n_i \rangle \simeq 0.385$, there is
a sharp drop in $n_s/m^\ast$, and $\langle n_i n_j \rangle$ begins
to increase from zero.  At this density the physical characteristics
of the system abruptly change, without however inducing a phase
transition.

\subsubsection{Partially-to-densely filled regime, $\langle n_i \rangle\gtrsim
0.385$}

For intermediate densities $\langle n_i \rangle \approx$
0.385--0.63, the kinetic energy $\langle K \rangle$ remains
approximately constant. Near $\langle n_i \rangle \simeq 0.63$,
there is a phase transition to a densely filled phase (either $D$ or
$A$). We go from a physical picture where the carriers are electrons
in a mostly empty background to one where the carriers are holes
moving in a mostly filled background. These holes condense into a
superfluid at lower temperatures, and the peak in $n_s/m^\ast$
occurs in the vicinity of the dilute-dense narrow first-order phase
transition. For $1/t \lesssim 0.16$, the maximum superfluid weight
is reached inside the $\tau$ phase. In the densely filled regime,
$\langle n_i \rangle \gtrsim 0.63$, the kinetic energy goes linearly
as $\langle K \rangle \simeq 2 (1-\langle n_i \rangle) = 2\delta$,
as expected if the free carriers are holes.

For hole-doped high-$T_c$ materials, the density of free carriers
increases with $\delta$ until the doping level optimal for
superconductivity is reached, and remains approximately constant in
the overdoped regime.\cite{Puchkov}  The superfluid weight, in
contrast, peaks near optimal doping and sharply decreases with
overdoping. These trends are reproduced in our numerical results,
identifying, from our calculated $n_s/m^\ast$ maxima, the optimal
doping for the $tJ$ model as $\delta \approx$ 0.32--0.37, the range
of densities where the $\tau$ phase occurs. Note that optimal doping
for high-$T_c$ materials is lower than this, typically around
$\delta = 0.15$, and the closely spaced double-peak structure of
$n_s/m^\ast$ at low temperatures near optimal doping is not
observed.  On the other hand, our approximation for the $d=3$ $tJ$
model is closer to experiment in this respect than earlier numerical
studies of the $tJ$ model, which focused mostly on finite-cluster
techniques applied to the $d=2$ system~\cite{DagottoRev}. In these
earlier studies optimal doping is identified near $\langle n_i
\rangle = 0.5$ on the basis of d-wave pairing correlations and the
peak in the superfluid weight~\cite{DagottoRiera}.  Also in these
earlier studies, the kinetic energy has a maximum at $\langle n_i
\rangle = 0.5$, but, unlike experiments, does not saturate with
overdoping~\cite{Dagotto2}.

To complete the description of the superfluid weight in this regime,
in Fig. 4 we show $n_s/m^\ast$ as a function of temperature $1/t$ at
various electron densities $\langle n_i \rangle$. For systems with
small to optimal hole dopings, shown in Fig. 4(a), there is a clear
onset temperature near $1/t \simeq 0.2$ below which the superfluid
weight rises rapidly, until it levels off near zero temperature.
This behavior is in good comparison with experimental results with
YBa$_2$Cu$_3$O$_{6+x}$~\cite{Lemberger}. As we move past optimal
doping to the overdoped systems of Fig. 4(b), we see a marked change
in behavior, with the low temperature $n_s/m^\ast$ suppressed.

\begin{figure} \centering

\includegraphics*[scale=1.0]{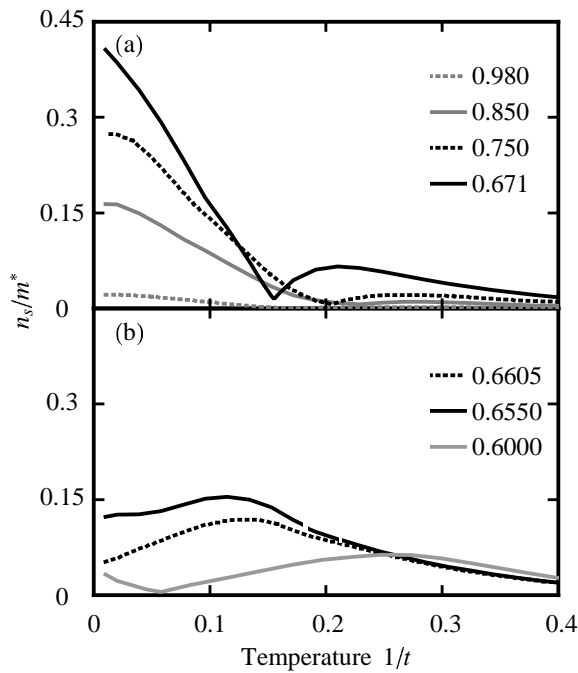}
\caption{The superfluid weight $n_s/m^\ast$ as a function of
temperature $1/t$ for various electron densities $\langle n_i
\rangle$ indicated in the legends.  Fig. 4(a) shows results in the
range of small to optimal hole doping, while Fig. 4(b) shows results
for hole overdoped systems.}
\end{figure}

\subsection{Specific Heat}

\begin{figure} \centering

\includegraphics*[scale=1.0]{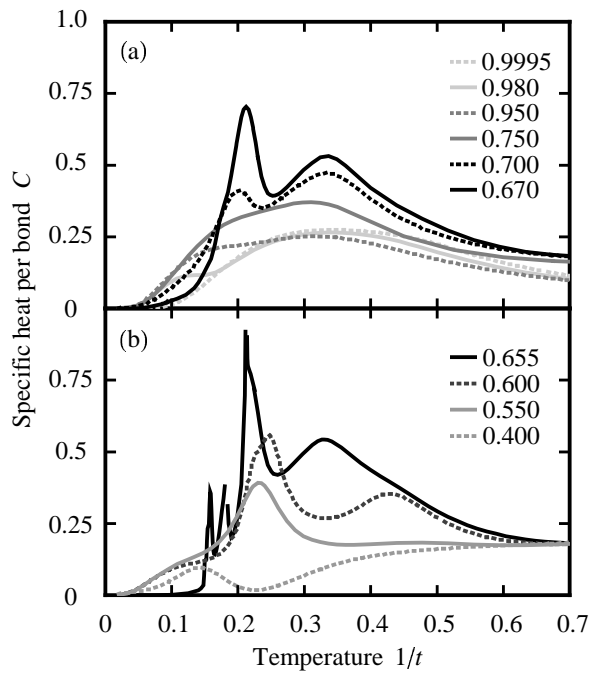}
\caption{The specific heat per bond $C$, in units of $k_B$, as a
function of temperature $1/t$ for various electron densities
$\langle n_i \rangle$ (indicated in the legends).  Fig. 5(a) shows
results in the range of small to optimal hole doping, while Fig.
5(b) shows results for hole overdoped systems.  The small
discontinuities in the plot for $\langle n_i \rangle = 0.655$
reflect temperature ranges where that particular density does not
appear because of the narrow first-order phase transition.}
\end{figure}

\begin{figure} \centering

\includegraphics*[scale=1.0]{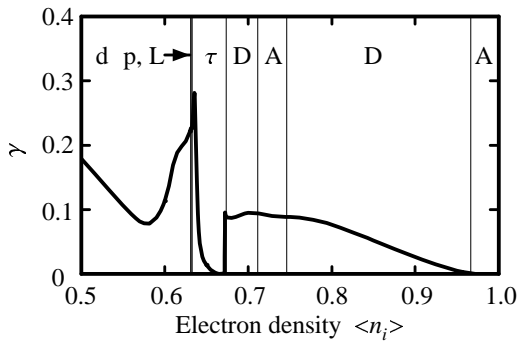}
\caption{The specific heat coefficient $\gamma = C/T$, in units of
$k_B^2$, as a function of electron density $\langle n_i \rangle$ at
temperature $1/t = 0.015$.  The corresponding phases are indicated
above the plot, and the location of phase boundaries marked by thin
vertical lines.  The symbol p refers to a region of forbidden
densities due to the narrow discontinuity at a first-order
transition. The symbol L refers to a ``lamellar'' region where
narrow slivers of the A and D phases alternate.}
\end{figure}

Since the superfluid weight peaks inside the $\tau$ phase at low
temperatures, it is interesting to check whether the $\tau$ region
has any other general characteristics of a superconducting phase. We
have added a magnetic field spin coupling term to the $tJ$
Hamiltonian and have shown that the $\tau$ phase continues to exist
when $H\ne 0$, up to a critical field $H_c(T)$, which decreases with
increasing temperature and goes to zero at the temperature of the
$\tau$ phase boundary. In our present study, we look at the spectrum
of excitations of the system through the specific heat per bond
\begin{equation}\label{eq:42}
C(n,T) = \left. \frac{\partial \langle H(i,j) \rangle}{\partial T}
\right|_n\,,
\end{equation}
calculated for $\phi =0$. If the $\tau$ phase corresponds to the
superconducting phase in real materials, we should see evidence of a
gap in the excitation spectrum.

The results for $C$ as a function of temperature $1/t$ are plotted
in Fig. 5 for a series of different electronic densities $\langle
n_i \rangle$.  Starting at $\langle n_i \rangle = 0.9995$, the
smallest hole doping shown in Fig. 5(a), we observe a broad peak
around $1/t \simeq 0.33$, corresponding to $k_B T \simeq 0.75
\tilde{J}$. We can identify this peak with the thermal excitation of
the spin degrees of freedom.  As we dope the system with holes, the
weight under the curve at lower temperatures increases due to
excitation of charge degrees of freedom.  As we approach optimal
doping, a second peak develops around $1/t \simeq 0.2$. Note that
this approximately coincides with the onset temperature below which
we see a dramatic increase in $n_s/m^\ast$ in Fig. 4. The
spin-excitation peak is also enhanced for $\langle n_i
\rangle\approx$ 0.65--0.75, which is related to the appearance of an
antiferromagnetic island around $1/t \simeq 0.3$ in that density
range.

The peak at $1/t \simeq 0.2$ grows rapidly near optimal doping,
reminiscent of the specific heat anomaly of high-$T_c$
materials~\cite{Loram1,Loram2}.  For $\langle n_i \rangle = 0.655$
we see the appearance of two subsidiary peaks below the main one at
$1/t \simeq 0.2$.  These smaller peaks may be related to the
complicated lamellar structure of A and D regions above the $\tau$
phase boundary.  For temperatures $1/t \lesssim 0.16$, inside the
$\tau$ phase, the specific heat is strongly suppressed, reflecting
the opening up of a gap in the excitation spectrum.  We can see this
gap more directly by looking at the low-temperature limit of the
specific heat.  Quasiparticle excitations contribute a linear term
to the specific heat $C \simeq \gamma T$ for small $T$.  In Fig. 6,
we plot $\gamma = C/T$ as a function of electron density at a low
temperature, $1/t = 0.015$. The specific heat coefficient $\gamma
\simeq 0$ in the A phase near half-filling, but then grows with
increasing hole doping.  At the onset of the $\tau$ phase a gap
opens in the quasiparticle spectrum, $\gamma$ falls sharply, and
stays small until it rises again near the phase boundary.
Qualitatively, this doping-dependence of the low-temperature
specific heat coefficient agrees well with the experimental results
for high-$T_c$ superconducting materials~\cite{Loram1}.\vspace{1em}

\section{Conclusions}

We have developed a position-space renormalization-group
approximation to study the superfluid weight of the
three-dimensional $tJ$ model. Our results indicate that optimal hole
doping for this system occurs in the density range of the $\tau$
phase, $\langle n_i \rangle \approx $ 0.63--0.68, where $n_s/m^\ast$
reaches a local maximum. While the superfluid weight drops off
sharply in the overdoped region, the density of free carriers,
proportional to the kinetic energy, remains approximately constant,
as seen experimentally in high-$T_c$ materials.  From calculations
of the specific heat coefficient $\gamma$, we see clear evidence of
a gap in the excitation spectrum for the $\tau$ phase. Earlier
renormalization group studies~\cite{FalicovBerkerT, FalicovBerker}
had suspected that the $\tau$ phase corresponds to the
superconducting phase of high-$T_c$ materials, and this idea was
reinforced when an analogous phase was discovered in the Hubbard
model~\cite{Hubshort,Hubnew}. Our present results justify this
suspicion.

\begin{acknowledgments}
This research was supported by the U.S. Department of Energy under
Grant No. DE-FG02-92ER-45473, by the Scientific and Technical
Research Council of Turkey (T\"UBITAK) and by the Academy of
Sciences of Turkey.  MH gratefully acknowledges the hospitality of
the Feza G\"ursey Research Institute and of the Physics Department
of Istanbul Technical University.
\end{acknowledgments}

\end{document}